# NON-DIMENSIONAL CONFINEMENT SCALING IN SIMILAR NEGATIVE TRIANGULARITY PLASMAS ON THE DIII-D AND TCV TOKAMAKS

[1]A. MARINONI, [2]C. CHRYSTAL, [3]S. CODA, [3]R. COOSEMANS, [1]C. MARINI, [3]M. PODESTA, [3]O. SAUTER, [4]M. AGOSTINI, [5]M.E. AUSTIN, [2]E. BELLI, [2]J. CANDY, [6]M. GORELENKOVA, [3]D. HAMM, [2]A.W. HYATT, [2]M. KNOLKER, [4]M. LA MATINA, [7]P. LUNIA, [8]S. MORDIJCK, [7]A.O. NELSON, [2]T.H. OSBORNE, [7]C. PAZ-SOLDAN, [3]L. PORTE, [3]U. SHEIKH, [9]F. SCOTTI, [2]K.E. THOME, [2]M. VAN ZEELAND and the DIII-D and TCV Teams.

[1] University of California San Diego, La Jolla CA, United States of America, Email: amarinoni@ucsd.edu
[2] General Atomics, San Diego CA, United States of America
[3] Ecole Polytechnique Federale de Lausanne, Lausanne, Switzerland
[4] Consorzio RFX, Padova, Italy
[5] University of Texas at Austin, Austin TX, United States of America
[6] Princeton Plasma Physics Laboratory, Princeton NJ, United States of America
[7] Columbia University, New York NY, United States of America
[8] College of William and Mary, Williamsburg VA, United States of America
[9] Lawrence Livermore National Laboratory, Livermore CA, United States of America

**Abstract**

Similarity experiments were performed on the DIII-D and TCV tokamaks to explore the scaling of energy confinement in negative triangularity plasmas using non-dimensional variables. Near up-down symmetric plasmas with large top-bottom averaged negative triangularity were created in a lower single null configuration, with the shape of the separatrix being closely matched between the two devices. The normalized energy confinement is found to weakly improve at increasing collisionality and, between the two devices, shows a machine size scaling behavior between Bohm and gyro-Bohm. Engineering scaling on a large DIII-D dataset is in agreement with the non-dimensional experiment.

## 1. INTRODUCTION

Negative triangularity (NT) configurations are emerging as potentially viable solutions for future reactors thanks to their ability to sustain H-mode grade levels of confinement and pressure despite the absence of an edge pedestal. The latter property is beneficial in several aspects, such as eliminating the need for active mitigation or suppression of edge localized modes (ELMs), low impurity retention, a wider Scrape-Off Layer heat flux width, and the possibility of employing impurity induced mantle radiation to dissipate power from inside the last closed flux surface [1]. Although results in present devices are encouraging, the design of a prospective reactor operating with NT shaping requires knowledge of how the energy confinement scales to reactor conditions. Indeed, the accuracy of predictions of fusion performance in future magnetic confinement devices heavily relies on that of the energy confinement time, whose value is determined by a combination of collisional and turbulent transport phenomena. Given the complexity of the physics at play, predictions from first principle models are extremely hard to obtain, especially in future devices where the lack of experimental information causes large uncertainties in the models. As a result, the energy confinement time is commonly extrapolated by employing regression formulas obtained by fitting its experimentally inferred values versus a number of engineering actuators that are believed to control the transport level in the plasma discharge. In order to avoid large uncertainties in extrapolating results to reactor conditions, or obtaining even wrong dependencies resulting from the Simpson paradox [2], the data-set must contain large enough variations in the independent parameters, which typically requires data from multiple devices. In the case of NT configurations, however, the paucity of devices that are able to create the desired plasma configuration severely curtails the variable space usable for the regression analysis. Besides uncertainties in the prediction resulting from errorbars associated with the regression variables, additional issues are associated with the identification of the correct independent variables to be considered, e.g. hidden or co-linear quantities, or the proper regression model [3]. As an example, the engineering confinement scaling in NT plasmas from the DIII-D armor campaign using automatic filtering of plasma parameters of interest was previously attempted [4]. An error analysis on the scaling coefficients resulted in extremely large uncertainties, possibly also due to the size of the dataset which consisted of only a few hundred points (for comparison, the IPB98(y,2) scaling dataset consists of over then thousands points spanning several devices). A different approach is offered by the method of non-dimensional analysis, or scale invariance, which consists of normalizing the equations governing a physical systems in such as way that those are expressed by a limited number of non-dimensional quantities that can be used to characterize the overall behavior of the system. The advantage of this approach is that the resulting non-dimensional quantities often have theoretical limits and offer a physical interpretation to the observed dependencies. Indeed, in contrast to the result from an engineering scaling, non-dimensional analysis can be compared in a much more straightforward way to numerical solvers of the equations governing the system; once the models recover the experimental measurements, they can be used to confidently extrapolate fusion performance to future reactors. While non-dimensional analysis was first developed at the beginning of the nineteenth century, it appears to have been applied to plasmas not before the 1960s [5,6] and specifically to tokamaks in the mid 1970s [7,8].

The manuscript is organized as follows. Section 2 overviews the basic principles of non-dimensional analysis, which guided the design of the experiments herein reported. Section 3 describes how the experiments were designed and executed. Section 4 and 5 report on the experiments on both devices using normalized gyro-radius and collisionality, respectively, as non-dimensional variables. A summary is offered at the end.

## 2. BASIC PRINCIPLES OF NON-DIMENSIONAL ANALYSIS

Ever since the 1990s, non-dimensional analysis has been used to design and execute experiments on a number of tokamaks worldwide to determine the scaling of the energy confinement time. An exhausting report of these experiments, as well as a description of the difference between non-dimensional and scale invariance techniques, can be found in a review paper from Luce et al. [9]. For the scope of this work it suffices to state that, in non-dimensional units, the confinement time normalized to the cyclotron frequency of a plasma species can be expressed as

$$\tau \Omega = f\left(\rho_*, \beta, \nu_*, \{r_i\}\right),$$

where f is a generic function not known a priori, $\tau$ is the confinement time, $\Omega$ the cyclotron frequency, $\rho_*$ the Larmor radius normalized to a convenient scale length, $\beta$ the plasma pressure normalized to the magnetic pressure, and $\nu_*$ the normalized collision frequency. The symbol $\{r_i\}$ stands for a set of ratio-like quantities entering the basic equations governing the system; these include the electron to ion temperature ratio, aspect ratio, poloidal to toroidal magnetic field intensity, effective charge, cross sectional shaping coefficients, or velocity normalized to the sound speed. It must be mentioned that this formulation holds only for plasma quantities and neglects the impact of atomic physics processes, which means that the penetration scale length of neutrals must be much smaller than the plasma size, a condition that is met in the experiments herein analyzed. Since the functional dependence of the normalized confinement time is not known a priori, non-dimensional experiments are most informative when they are executed in such a way as to vary only one parameter at a time while keeping everything else fixed. The functional dependence f is reconstructed as a power law consisting of a single term whose exponent is determined by a numerical fit to experimental data. In the case of collisionality dependence, this can be

$$\tau \Omega = A\, \nu_*^{\alpha},$$

with $\alpha$ being the scaling parameter to be determined and A representing the contribution from the other non-dimensional variables. The non-dimensional variables can be modified by varying either the plasma size or discharge parameters because of their mutual relations

$$q \propto B_T a \varepsilon / I_P; \qquad \beta \propto nT/B_T^2; \qquad \rho_* \propto T^{1/2}/(B_T a); \qquad \upsilon_* \propto \varepsilon^{-3/2} nqR/T^2;$$

where T is the plasma temperature, $\varepsilon$ the inverse aspect ratio, a the plasma minor radius, R the major radius, $I_P$ the plasma current, n the plasma density, $B_T$ the confining magnetic field, and q the safety factor. Hence, variations in one non-dimensional quantity with others held fixed can be obtained by appropriately adjusting discharge quantities, both in experiments and in numerical models. We note that, for a given value of the inverse aspect ratio, the safety factor in circular plasmas yields the ratio of the like quantities poloidal and toroidal fields. It has to be stressed, however, the this is no longer true when comparing plasmas featuring different poloidal cross sectional shapes because, in addition to the poloidal to toroidal field ratio, the safety factor also depends on the plasma shape itself. It is therefore convenient to execute experiments at fixed plasma shape for the relations above to hold their role in interpreting results. Table 1 displays, for a fixed aspect ratio of the device, the required modifications to the discharge quantities in order to carry out scans in one non-dimensional variable while maintaining the others fixed.

TABLE 1.   RELATIONS AMONG PLASMA DISCHARGE QUANTITIES RESULTING IN A SINGLE NON-DIMENSIONAL QUANTITY SCAN

| | | | |
|---|---|---|---|
| $\rho_*$scan: $\rho_* \propto B_T^{-2/3} a^{-5/6}$ | $I_P \propto B_T a$ | $n \propto B_T^{4/3} a^{-1/3}$ | $T \propto B_T^{2/3} a^{1/3}$ |
| $\upsilon_*$ scan: $\nu_* \propto B_T^{-4} a^{-5}$ | $I_P \propto B_T a$ | $n \propto a^{-2}$ | $T \propto B_T^2 a^2$ |
| $\beta$ scan: $\beta \propto B_T^4 a^5$ | $I_P \propto B_T a$ | $n \propto B_T^4 a^3$ | $T \propto B_T^2 a^2$ |

### 3. DESCRIPTION OF THE EXPERIMENTS

While non-dimensional experiments have been carried out in the past on various devices in both H-mode and L-mode plasmas (see ref. [9] for a review), this work reports on the first sedulous experimental investigation of NT confinement scaling across multiple machines, attempting to determine the dependence of the energy confinement on non-dimensional units. Experiments were performed on the DIII-D and the TCV tokamaks, which were chosen because they are the only two devices able to realize strongly shaped NT plasmas, for which the impact of triangularity would appear more clearly in the data. Near up-down symmetric plasmas with large top-bottom averaged triangularity, $\delta_{avg} \sim$ -0.5, were created in a lower single null configuration, displayed in Fig.1, with the shape of the separatrix being closely matched between the two devices. Differences in the position of the separatrix seen in Fig. 1 are generally consistent with uncertainties in the determination of the separatrix in the two devices. The TCV equilibrium was created by subtracting the position of the LCFS centroid from the DIII-D equilibrium, scaling dimensions by the ratio of the minor radii of the two devices, and adding the desired centroid position for TCV. This results in inverse aspect ratio equal to $\varepsilon \approx 0.32$ for DIII-D and $\varepsilon \approx 0.28$ for TCV.

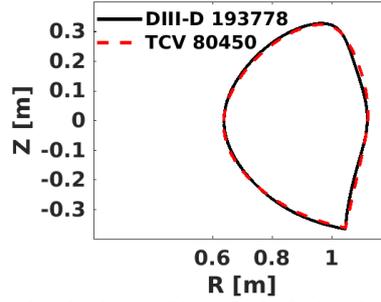

*FIG. 1.Comparison of the separatrix for TCV discharge #80450 (dashed red) and DIII-D discharge #193778 (full black) scaled to fit inside the TCV vessel. Both discharges held steady equilibria during the plasma current flat-top phase.*

Experiments were designed with the primary goal of executing non-dimensional scaling studies of confinement on the two devices separately. Additionally, a $\rho_*$ scan across the two devices was also attempted in an effort to extend the operational space over which this non-dimensional quantity was varied with all others kept fixed. Given the resources made available to both experiments, it was decided to prioritize scans in $\rho_*$ and $\upsilon_*$ over $\beta$ and ratios of other like quantities. This was dictated both by the importance of $\rho_*$ and $\upsilon_*$ in extrapolating energy confinement values to reactor conditions, and by practical reasons. For example, since the on-axis safey factor is usually maintained near unity by saw-teeth, a scan in the safety factor entails a variation in the magnetic shear as well, which severely complicates the interpretation. In the case of a $\beta$ scan on a single device, as it is apparent from Table 1, it is necessary to vary the plasma density by the same factor by which $\beta$ is varied. It is generally problematic to run discharges with largely different plasma density but small differences in plasma current, and it might cause variations in the radial dependence of the coupled NBI power. Additionally, depending on the minimum operational density of the device, a large enough variation of $\beta$ might result in the plasma to either operate in a detached divertor condition, or even attain the density limit. Since both of these conditions are seen to significantly affect the confinement by altering the plasma profiles near the edge, the interpretation of results would be severely complicated with a different plasma boundary rather than being the result of a core confinement scaling. Given operational difficulties in the DIII-D $\rho_*$ scaling, which will be described in Sec. 3, it was decided to not attempt the $\beta$ scaling experiment for lack of experimental time available and to try finalizing the $\rho_*$ scaling part of the experiment. Correspondingly, the TCV experiment did not attempt the $\beta$ scaling because it would not have had a comparison point from DIII-D.

On both devices, magnetic field, auxiliary power, plasma current and electron density were altered in such a way as to individually vary only the non-dimensional variable being studied, while maintaining all the other quantities fixed over most of the normalized minor radius. To offer a quantitative view of the effect of the scaling procedure displayed in Table 1, the radial profiles of the plasma electron density, temperature, ion temperature and toroidal velocity for DIII-D data are displayed in Fig. 2 before and after the scaling procedure is implemented. Experiments were designed to explore primarily beam-heated plasmas without radio-frequency (RF) heating to avoid inconsistencies when exploring low-field discharges, for which RF auxiliary power would not be effectively coupled. This resulted in plasmas with electron to ion temperature ratio near unity and effective charge below and above two, for DIII-D and TCV respectively, throughout most of the plasma volume. On both devices, NT plasmas routinely operate with the highest confinement at edge safety factor less than three, and the energy confinement is observed to strongly improve with increasing plasma current [10,11]. As such, all discharges on both devices were executed near the lowest achievable edge safety factor which, owing to the machine operational constraints of the joint experiment, corresponds to $q_{95} = 2.8$. The value of the safety factor on axis was maintained near unity by saw-teeth, thereby making the magnetic shear over radius as similar as experimentally achievable among all cases. No active techniques to fine tune the shape of the safety factor, such as radially localized RF current drive, was used. In order to obtain MHD quiescent discharges that would be suitable for transport analysis, plasmas at low auxiliary torque were not explored because they are generally more prone to the appearance of MHD activity, especially in discharges with magnetic field lower than the standard value typically used on the device.

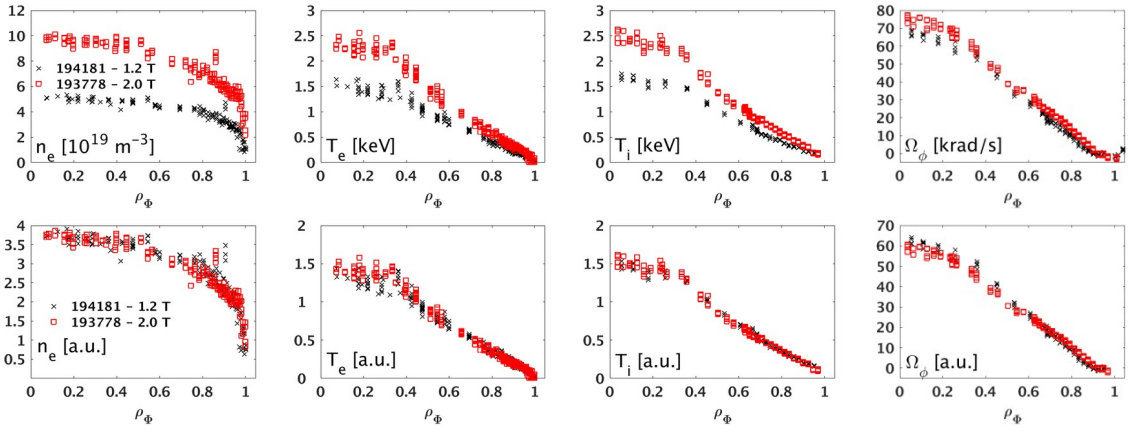

*FIG. 2. Comparison of radial profiles for a $\rho_*$ scan on DIII-D in physical units (top) and after the corresponding normalization described in Table 1 (bottom). From left to right: electron density, electron temperature, ion temperature, ion toroidal rotation.*

Once the magnetic field and the plasma current were set to the desired values, fine increments in auxiliary power were implemented while maintaining a constant line averaged density using active feedback control on the gas injection valve. The time evolution of the auxiliary power was incremented as a staircase, with steps longer than a few energy confinement and fast ions thermalization times, so as to ensure that profiles relax to a stationary state. Experiments on both devices used the following procedure: the non-dimensional quantity of interest was varied by modifying the confining magnetic field, with values for the plasma current, line averaged density, and auxiliary power adjusted so that the radial profiles of plasma parameters of interest would scale according to the relations in Table 1. After a given discharge was executed, it was verified that the measured radial profiles of plasma electron density, electron and ion thermal temperatures, and plasma toroidal rotation varied by the desired amount. In case the profiles did not vary as expected, the line integrated density was adjusted, and the power scan repeated in a subsequent discharge until the desired result was obtained. For all cases on both devices, the energy confinement time is obtained from a power balance analysis using the TRANSP code [12], and accounting for the total radiated power from the plasma core. The impact of impurity content and fast ions diffusion on the energy confinement time and diffusivities was evaluated in the simulations: the effective charge content was varied between the values measured by the charge exchange recombination spectroscopy systems and that obtained by matching the measured loop voltage, while fast-ions diffusion was artificially incremented to recover the measured neutron rate. While numerical values of the energy confinement time are affected by these variations, its non-dimensional scaling is impacted to a much lower extent as long as it is computed using simulations with similar settings. Although the beam power deposition generally varies at varying plasma density, the density variation employed in this experiment is not large enough to cause significant differences in the beam power deposition: for the cases reported in Fig. 2, the cumulative integral within 50% of the plasma volume, a value taken for illustration purposes, is computed to be equal to 89% and 92% of the total for the high and low field cases, respectively. Typical uncertainties on local plasma values, e.g. normalized collisionality at mid volume, are estimated as population uncertainties within 10%-15%; these values are omitted below to avoid heavy notation.

## 4. RESULTS FROM THE NORMALIZED LARMOR RADIUS EXPERIMENT

In terms of amount of run-time, most of the effort on both devices was devoted to the $\rho_*$ experiment because of its importance in projecting performance to future devices which, presumably, will be characterized by large size and confining magnetic fields of high intensity. The importance of the normalized gyro-radius scaling can be understood by considering that no present device can operate in the $\rho_*$ parameter space that reactors will work at, while that of the other non-dimensional variables is accessible. For the reader unfamiliar with the subject, it can be shown that the non-dimensional confinement scaling can be cast in the form $B_T\tau \propto \rho_*^{-2-\alpha}$; the cases corresponding to $\alpha = 0$ and $\alpha = 1$ are referred to Bohm and gyro-Bohm confinement, respectively, in the literature. Two separate scans were performed on the two devices, as well as an attempt to a $\rho_*$ cross machine comparison.

### 4.1. Normalized Larmor radius scan on DIII-D

In view of the fact that, when keeping other non-dimensional quantities fixed, the normalized Larmor radius scales weakly with the confining magnetic field as $\rho_* \propto B_T^{-2/3}$, the magnetic field must be varied by approximately a factor of two to obtain a 60% variation in $\rho_*$, a quantity that is barely large enough for a credible regression. While achieving the plasma breakdown at low magnetic field did not pose much problem, it was extremely difficult to obtain matched profiles over most of the minor radius when comparing discharges at full field, i.e. $B_T = 2$ T, to those at field reduced below $B_T = 1.6$ T. In particular, as displayed in Fig. 3, the properly scaled radial profile of the electron density shows much higher peaking factor at low field ($B_T = 1.2$ T) as compared to the high field case ($B_T = 2$ T), even though the scaled line averaged density has the desired value. This behavior persisted despite efforts were made in modifying the discharge trajectory (e.g. time evolution of the plasma current, beam heating or torque).

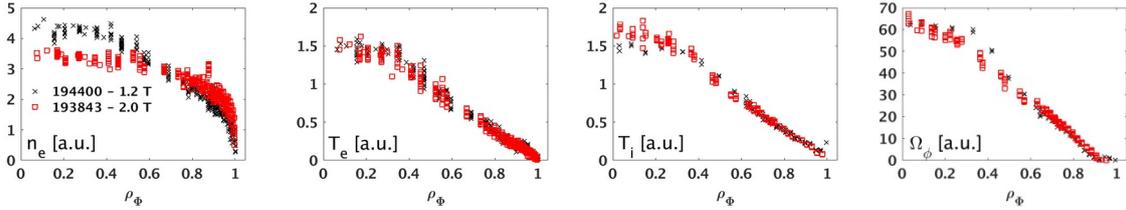

*FIG. 3. Radial profiles of the plasma electron density, electron temperature, ion temperature and toroidal rotation as a function of the normalized toroidal flux on DIII-D, scaled for a $\rho_*$ scan.*

Clearly, unmatched profiles in scaled units correspond to variations in other non-dimensional quantity, thereby invalidating any conclusions based on a regression analysis that assumes $\rho_*$ as the only quantity being varied. Nonetheless, a survey of all discharges at full field was carried out to identify the closest matches to the low field discharges in the database. When the regression was carried out between the closest pairs, the $\rho_*$ dependence of the normalized confinement time was found to be of the form $B_T\tau \propto \rho_*^{-2+\alpha}$ with $\alpha > 0$, corresponding to Bohm scaling of confinement at best [13].

A reasonable match of all plasma scaled profiles was obtained only between a $B_T = 2.0$ T case in stationary phase and a $B_T = 1.2$ T case in a phase during which the line averaged density was quickly increased via high throughput gas injection; the difference in magnetic field between these two cases corresponds to a 40% variation in $\rho_*$. Unfortunately, the low field discharge could not be repeated in such a way as to make it stationary in the phase of interest due to lack of time devoted to this experiment. The two discharges feature $\beta_N \sim 2.2$ and mid-volume collisionalities $\nu_{*e} \sim 0.5$ and $\nu_{*i} \sim 0.3$. The radial plasma profiles scaled according to the $\rho_*$ scaling in Table 1 are displayed in the lower row of Fig. 2 , featuring excellent agreement over almost the entirety of the plasma volume. The time dependent transport analysis carried out with TRANSP includes the impact of the non stationary condition of the low field case, and indicates that the normalized energy confinement time scales with the normalized Larmor radius as $B_T\tau \propto \rho_*^{-2 +/- 0.7}$, where the large uncertainty in the exponent is mostly due to the time varying conditions of the low field discharge, which is interpreted as Bom scaling of confinement. Further inspection of the transport results reveals that the Bohm scaling of confinement is mostly due to the ion transport, while electrons feature gyro-Bohm scaling over most of the plasma volume. The effective one fluid energy diffusivity shows a profile intermediate between those of the two fluid species, as expected. This is displayed in Fig. 4, where the $\rho_*$ scaling is expressed as energy diffusivity through the relation $\tau \approx \chi/a^2$, where a is the plasma minor radius and $\chi$ the energy diffusivity. A similar conclusion regarding the difference between ion and electron transport was obtained independently in the survey of near-matched discharges [13]. Such behavior, which is consistent with previous experiments in L-mode plasmas carried out on the DIII-D [14], JT-60U [15] and Tore Supra [16] tokamaks, can perhaps be loosely understood in terms of relative size of the particles' Larmor radius relative to the scale length of the profile driving the turbulent transport, a quantity that is much larger for ions, when the energy flux in the ion channel is a significant fraction of the total.

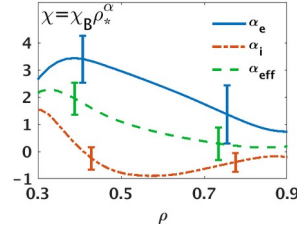

*FIG 4. Radial profiles of the inferred gyro-radius scaling for the energy diffusivity for electrons (blue full), ions (red dash-dotted red), and one fluid effective (green dashed) for the DIII-D $\rho*$ scan reported in Fig. 2.*

It is interesting to evaluate the engineering scaling of thermal energy confinement time over all the discharges carried out on DIII-D which have a shape similar to that displayed in Fig. 1. All discharges were manually inspected to select suitable time windows that are at least three energy confinement times long and during which plasmas are stationary in the actuators, stored energy, as well as in the radial profiles of density, temperature, and velocity. In view of the fact that the database contains discharges from various experiments, each of them having different discharge trajectories and utilizing different actuators being pursuant of different goals,  the selected set of discharges was further reduced by subsequently applying additional filtering on the effective charge, the ion to electron temperature ratio, presence of MHD activity, proximity to the Greenwald limit, and presence of impurities that were used to probe divertor detachment in dedicated sessions. Regression over the plasma current, confining magnetic field, line averaged density and auxiliary power was carried out, resulting in a fit that can be cast in the form

$$\tau \propto I_p^{1.16} B_T^{0.16} n_e^{0.19} P_{aux}^{-0.75},$$

with adjusted coefficient of determination $R^2_{adj} = 0.91$. This, when converted into non-dimensional variables, yields

$$B_T \tau \propto \rho_*^{-1.97} \upsilon_*^{-0.05} \beta^{-2.25} q^{-4.81},$$

whose normalized gyro-radius and collisionality dependencies substantially agree with those obtained from the controlled non-dimensional experiment. It has to be stressed that confidence in the validity of these results is provided by the fact that the engineering and the non-dimensional experiments are consistent with each other both in the normalized Larmor radius and, as it will be shown in Sec. 5.1, normalized collisionality scans. When comparing to the ITER97-L scaling law developed for positive triangularity L-mode plasmas ($\tau \propto I_p^{0.96} B_T^{0.03} n_e^{0.40} P_{aux}^{-0.73}$), apart from the constant factor in front of the equation that is larger for NT plasmas, the difference is in a less pronounced dependence on plasma density and slightly stronger dependency on plasma current, as well as its near co-linear variable magnetic field, while the power degradation of thermal confinement is similar. In non-dimensional units, the ITER97-L scaling law displays Bohm confinement, improvement with normalized collisionality, and weaker dependencies on safety factor and plasma normalized pressure ($B_T \tau \propto \rho_*^{-1.85} \upsilon_*^{0.21} \beta^{-1.41} q^{-3.56}$).

### 4.1. Normalized Larmor radius scan on TCV

The confining magnetic field was varied between $B_T = 0.8$ T and $B_T = 1.44$ T, resulting in a 50% variation in $\rho*$. When using the $\rho*$ scaling factors formulas in Table 1, the scaled profiles display a good match over most of the plasma volume, as displayed in Fig. 5. Discharges feature $\beta_N \sim 1.2$ and mid-volume collisionalities $\nu*_e \sim 1.5$ and $\nu*_i \sim 1.2$, while the minimum collisionality values are reached around $\rho_\phi \sim 0.4$ and are approximately equal to $\nu*_e \sim 0.6$ and $\nu*_i \sim 0.5$. The regression analysis yields near gyro-Bohm scaling both for the normalized energy confinement, as well as for the ion and electron energy diffusivities. The reason for the observed Bohm vs gyro-Bohm scaling in the two devices might be due to the different normalized pressure, or collisionality, the two experiments operated at. In particular, it is noted that while DIII-D discharges tend to operate with effective charge in the range $1.5 < Zeff < 1.8$, while TCV discharges generally feature higher impurity content in the range $2.1 < Zeff < 2.6$, resulting in a more pronounced main ion dilution. A detailed study aimed at interpreting these results will be the subject of a future publication.

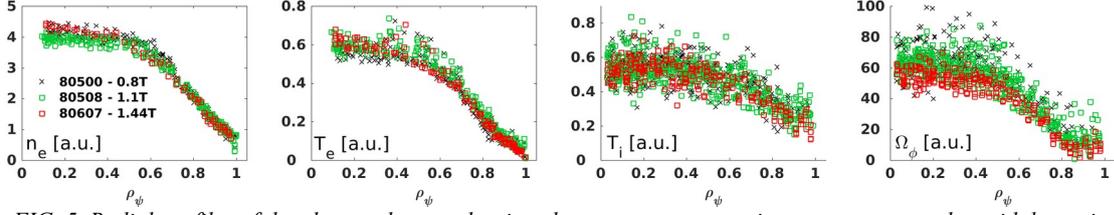

*FIG. 5. Radial profiles of the plasma electron density, electron temperature, ion temperature and toroidal rotation as a function of the normalized toroidal flux for the $\rho_*$ scan on TCV.*

A remarkable difference with respect to the DIII-D experiment is the fact that the plasma density profile shows very similar peaking factor regardless of the magnitude of the confining magnetic field. However, the density peaking on TCV is significantly higher than that on DIII-D, even for the low field cases. In order to carry out a cross machine $\rho_*$ scaling, dedicated discharges on TCV at high field ($B_T = 1.44$ T) were executed with the goal of matching scaled profiles with low field cases from DIII-D. If matched profiles at the high TCV field were obtained, low field discharges would be executed to further extend the $\rho_*$ variation. Additional third harmonic ECH heating was coupled to the plasmas in an effort to reduce the density profile peaking factor. Unfortunately, while a satisfactory agreement was found on the electron and ion temperature radial profiles, TCV plasmas are found to rotate approximately 20% faster on-axis and feature higher density peaking as compared to DIII-D plasmas, thereby preventing the determination of the cross-device $\rho_*$ scaling.

## 5. RESULTS FROM THE COLLISIONALITY EXPERIMENT

Two separate normalized collisionality scans were executed on the two devices which, as indicated by Table 1, consisted of varying magnetic field, plasma current and auxiliary power at fixed plasma density.

### 5.1. Normalized collisionality scan on DIII-D

Figure 6 displays the radial profiles of electron density, electron temperature, ion temperature and toroidal rotation for two stationary cases. Profiles from the low magnetic field discharge are normalized by the quantities obtained from Table 1 that guarantee that other non-dimensional quantities are held fixed.

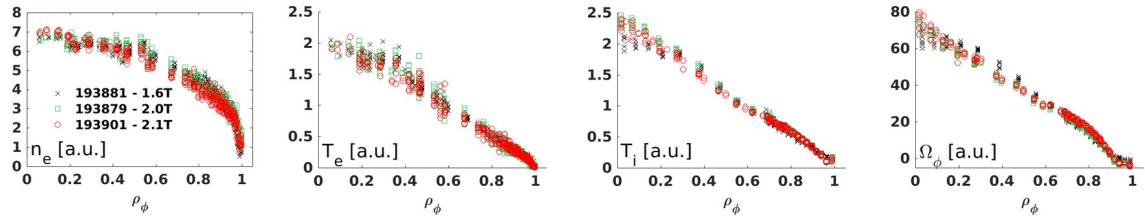

*FIG. 6. Radial profiles of the normalized plasma electron density, electron temperature, ion temperature and toroidal rotation as a function of the normalized toroidal flux on DIII-D. For visual clarity, only data collected over a 50 ms time window are included in each radial profile.*

Discharges on DIII-D were executed at $B_T = 2.1$ T and $B_T = 1.6$ T, resulting in the normalized collisionality being varied by a factor 2.9. Discharges feature $\beta_N \sim 2.0$, $0.14 < \nu_{*e} < 0.39$ and $0.08 < \nu_{*i} < 0.22$ near mid volume, while the minimum normalized collisionality values are reached near $\rho = 0.4$ and are $0.05 < \nu_{*e} < 0.15$ and $0.03 < \nu_{*i} < 0.11$. The regression yields a near collision-less dependence of the non-dimensional confinement $B_T\tau \propto \nu_*^{0.04 \, +/- \, 0.05}$. The uncertainty on the exponent was obtained using a Monte-Carlo analysis of the uncertainties associated to the quantities being regressed, without including the impact of partially unmatched non-dimensional quantities. The values of collisionality from the $\rho_*$ scan in Sec. 4 are at the higher limit of the collisionality scan explored. The collisionless behavior suggests that, at least to first order approximation, the $\rho_*$ result should hold even at lower collisionality than that at which it was obtained.

### 5.1. Normalized collisionality scan on TCV

Discharges on TCV were obtained using a similar procedure to that used on DIII-D. The confining magnetic field was varied between $B_T = 1.1$ T and $B_T = 1.44$ T, resulting in a 2.94 variation in normalized collisionality. Discharges feature $\beta_N \sim 1.15$, $0.28 < \nu_{*e} < 0.74$ and $0.42 < \nu_{*i} < 1.24$ near mid-volume. The normalized profiles are displayed in Fig. 7, and the regression yields $B_T\tau \propto \nu_*^{0.09 \, +/- \, 0.03}$.

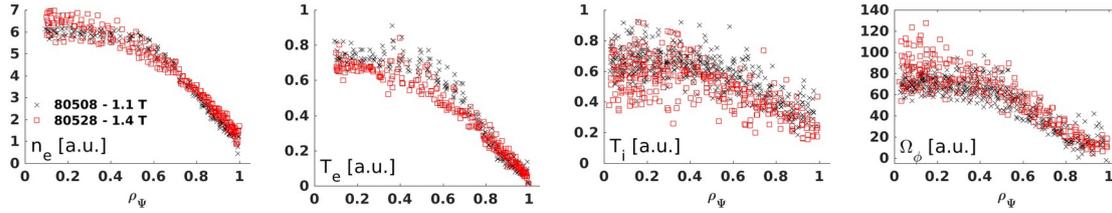

*FIG. 7. Radial profiles of the plasma electron density, electron temperature, ion temperature and toroidal rotation as a function of the normalized toroidal flux for the $\nu_*$ scan on TCV.*

The mis-match in the electron temperature near mid-radius is noted. If more power is added to the high field case to more closely match the electron temperature profile, a disagreement on the ion data is introduced. Although such mismatch is not expected to greatly affect the results, dedicated modelling should be used to quantify its actual effect. The collisionality on TCV is higher than that on DIII-D and reaches the collisional regime near mid-volume, although it is collisionless at inner radii; this is consistent with the larger positive collisionality exponent than that obtained for the corresponding experiment on DIII-D because of the fact that increasing collisionality is generally stabilizing, resulting in better confinement. In this respect, it is interesting to note that prior non-linear gyro-kinetic modelling work [17] aimed at predicting the non-dimensional confinement scaling in NT plasmas utilizing DIII-D inner-wall limited discharges from 2017 experiments. Those discharges, due to a combination of higher effective charge and density, were characterized by larger collisionality than the experiments here reported. Those simulations predicted a more pronounced dependence on collisionality, $B_T\tau \propto \nu_*^{0.5+/-0.2}$, which is consistent with collisionality quenching of micro-intabilities.

Finally, it has to be noted how the results from the normalized collisionality scan on both devices are consistent with previous experiments at positive triangularity in L-mode plasmas, for which the non-dimensional energy confinement is weakly collisional [9].

6. SUMMARY

The first non-dimensional confinement experiments in self similar shape were carried out for negative triangularity plasmas on the TCV and DIII-D tokamaks. Lower single null discharges were executed on both devices with closely matched separatrix shapes. Experiments were realized using auxiliary beam heating, $q_{95} = 2.8$, and electron to ion temperature ratio near unity over most of the plasma volume. Non-dimensional scans investigating the normalized gyro-radius and collisionality scalings were executed on both devices. Scaled radial profiles of plasma electron density and temperature, ion temperature and toroidal rotation were generally matched with very good accuracy across the minor radius. The gyro-radius scan on DIII-D shows near Bohm scaling for the confinement time, consistent with the engineering scaling carried out on a much larger dataset. The Bohm scaling is determined by the ion behavior, while electron energy diffusivity scales more favorably. The TCV experiment shows gyro-Bohm confinement for both ions and electrons. The difference between the $\rho_*$ scalings on the two devices might be due to the difference in normalized pressure and/or collisionality. The collisionality part of the experiment shows a weak, though positive, dependence on collisionality on both devices, which is understood in terms of collisionality quenching of micro-instabilities. The results herein reported offer a quantitative path to extrapolate the energy confinement time in NT configurations to future devices, as well as a way to interpret the basic principles governing transport in NT plasmas. While uncertainties remain in the extrapolation of NT performance in future devices due to the finite extent of the scans here reported, the present experiment can, and should, be used to validate first principle transport solvers which can then be utilized to improve confidence in the extrapolation of performance to future reactors.

**ACKNOWLEDGEMENTS**

This material is based upon work supported by the U.S. Department of Energy, Office of Science, Office of Fusion Energy Sciences, using the DIII-D National Fusion Facility, a DOE Office of Science user facility, under Awards DE-FC02-04ER54698, DE-AC02-05CH11231, DE-FG02-97ER54415, DE-SC0019302, DE-SC0022270, DE-FG02-95ER54309, DE-FG02-07ER54917. The work is also supported in part by the Swiss National Science Foundation. This work uses the TRANSP code, which is funded by Princeton Plasma Physics Laboratory / Princeton University under contract number DE-AC02-09CH11466 with the U.S. Department of Energy. The United States Government retains and the publisher, by accepting the article/presentation for publication, acknowledges that the United States Government retains a non-exclusive, paid-up, irrevocable, world-wide license to publish or reproduce the published form of this manuscript, or allow others to do so, for United States Government purpose.